# Laterally Excited Bulk Acoustic Wave (LBAW) X-Cut Lithium Niobate Resonators

Walter Gubinelli, Ryan Tetro, Pietro Simeoni, Luca Colombo, and Matteo Rinaldi
Institute for NanoSystems Innovation (NanoSI), Northeastern University, Boston, MA, USA

*Abstract*—In this work, Laterally excited Bulk Acoustic Wave (LBAW) resonators on X-cut Lithium Niobate ($LiNbO_3$) and, for the first time their higher-order overtones (LOBAW) are demonstrated by embedding interdigitated electrodes recessed into the piezoelectric thin film, allowing to exploit both $S_0$ and $SH_0$ vibrational modes. This recessed electrode architecture decouples the dispersion relation from film thickness, enabling lithographic tuning of resonance frequency and on-chip multi-frequency scaling on a single substrate, while concurrently increasing static capacitance density (**$C_0$**) and reducing ohmic losses (**$R_S$**). The excited $SH_0$ modes exhibits Figures of Merit (FoM) of 437 at 673 MHz for the fundamental tone and 53 at 1.05 GHz for the overtone. The proposed architecture holds large potential for future 5G/6G advanced radio frequency front-end modules, enabling on-chip multi-frequency scaling and improved performance.

## I. Introduction

THE unprecedented growth in Internet of Things (IoT) ecosystems and the rapid evolution towards Artificial Intelligence (AI)-driven applications have significantly expanded the landscape of wireless communications [1] [2]. The resultant proliferation of interconnected devices across various sectors, such as smart homes, autonomous vehicles, healthcare monitoring, and industrial automation, has heightened the demand for advanced Radio Frequency Front-End (RFFE) modules and integrated sensor nodes [3] [4] [5]. These applications necessitate robust and highly adaptable wireless communication technologies capable of meeting stringent requirements for bandwidth, latency, and power efficiency [6]. The ongoing expansion into higher frequency bands, crucial for next-generation wireless networks to be adopted for 5G and 6G applications, such as the sub-6 GHz Frequency Range 1 (FR-1), accentuates the need for scalable, tunable, and reconfigurable platforms in the millimeter-wave range [3] [7].

Piezoelectric Microelectromechanical Systems (MEMS) acoustic resonators represent a promising solution to address these emerging challenges, providing overall good performances, compact form factor, scalability, and cost-effective manufacturability [8]. Their compatibility with complementary metal–oxide–semiconductor (CMOS) integration further enhances their appeal as candidates for monolithically integrated, advanced RF front-end devices, enabling the deployment of efficient and reliable high-frequency protocols in wireless communication ecosystems [8] [9].

This work explores a novel alternative to traditional technologies commonly employed in RFFE architectures, i.e. Surface Acoustic Wave (SAW), Bulk Acoustic Wave (BAW), and Lamb Wave resonators. Efficient excitation of Lateral Bulk Acoustic Waves (LBAW) and, for the first time, their higher-order overtones (LOBAW) is hereby demonstrated by leveraging recessed electrodes in thin-film Lithium Niobate ($LiNbO_3$). As detailed in the following sections, LBAW resonators hold potential to overcome several limitations traditionally associated with MEMS acoustic devices. The first section compares traditional Lamb Wave resonators (LVRs) with the proposed LBAW technology, highlighting key performance metrics and associated technical challenges. Next, the multiphysics-based modeling approach for resonator design and optimization is presented. The subsequent section outlines fabrication methods and experimental results. Finally, the conclusions discuss potential strategies to further enhance device performance.

## II. Background and Motivation

Surface Acoustic Wave (SAW) resonators, historically first to be employed in RF filtering, exploit the propagation of Rayleigh acoustic waves along the surface of piezoelectric substrates [10]. However, their performance is naturally limited, especially at higher frequencies (2-3 GHz). SAW devices usually exhibit low electromechanical coupling ($\approx 5\%$); additionally, when scaled to the GHz range, the increased acoustic and electrical losses further degrade their quality factor and power handling capability [11]. Improved variants, such as Incredible High Performance SAW (IHPSAW), have been developed utilizing multilayered substrates, such as Lithium Tantalate LiTaO$_3$ on SiO$_2$/AlN Bragg reflectors. By enhancing the energy confinement in the acoustic cavity it is possible to mitigate losses, as well as improving the electromechanical coupling ($\approx 10\%$). However, IHPSAW resonators still fail to meet higher frequency requirements (typically above 5 GHz) [12] [13].

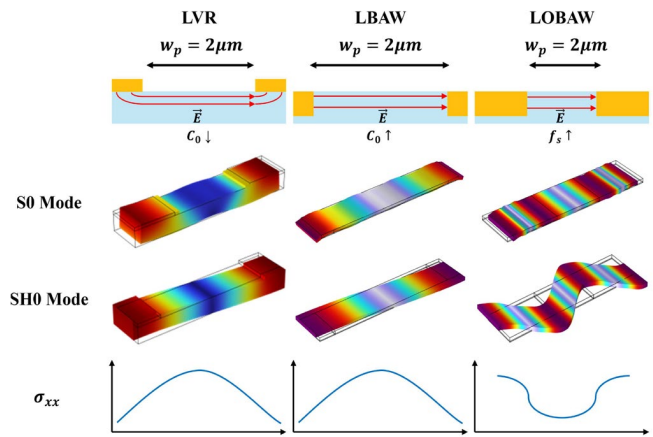

Fig. 1. Schematic comparison of laterally excited resonator configurations (LVR, LBAW and LOBAW) with a finger pitch $w_p = 2\mu m$, showing top-view electrode layouts and electric field distribution, COMSOL® Multiphysics simulated $S_0$ and $SH_0$ mode shapes, and $\sigma_{xx}$ stress distributions along the plate midline.

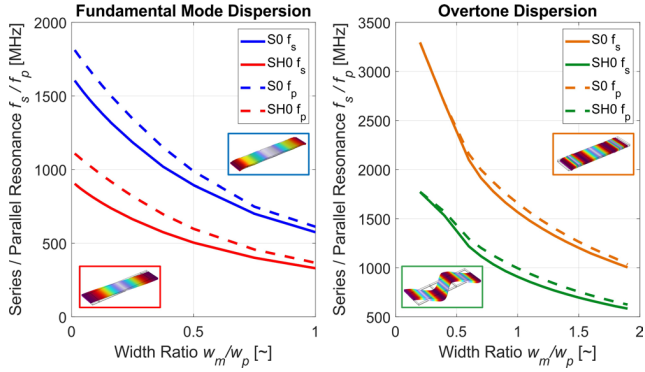

Fig. 2. COMSOL® Multiphysics simulated dispersion curves for resonance and antiresonance frequencies as function of $w_m/w_p$ ratio for fundamental $S_0$ and $SH_0$ modes (left) and their overtones (right).

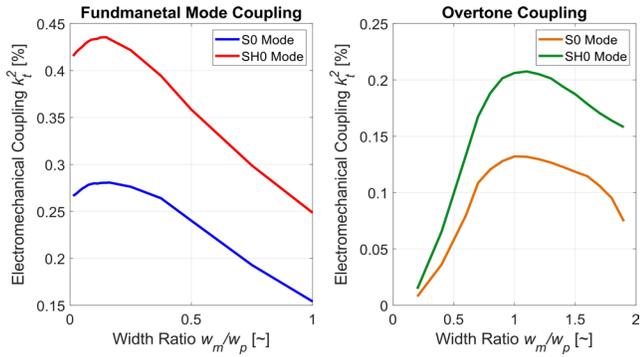

Fig. 3. COMSOL® Multiphysics simulated dispersion curves for electromechanical coupling as function of $w_m/w_p$ ratio for fundamental $S_0$ and $SH_0$ modes (left) and their overtones (right).

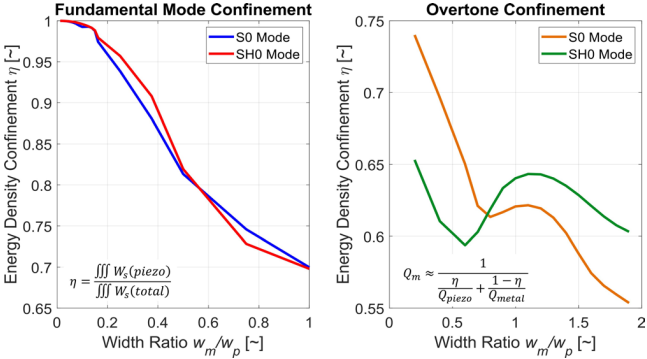

Fig. 4. COMSOL® Multiphysics simulated dispersion curves for energy confinement ratio $\eta$ as function of $w_m/w_p$ ratio for fundamental $S_0$ and $SH_0$ modes (left) and their overtones (right).

Thin Film Bulk Acoustic Resonators (FBARs) and their advanced variant, Periodically Poled Piezoelectric Film (P3F) resonators, provide a notable alternative to SAW devices [14] [15]. BAW resonators leverage thickness-extensional vibrations in thin piezoelectric films, offering significant advantages in compactness and higher operational frequencies (up to 10 GHz) compared to SAW devices, as well as increased Figure of Merit [16]. However, the fabrication of BAW devices demands stringent process control and is significantly more complex than that of conventional SAW devices; moreover, their resonance frequency is primarily set by the layer stack, making multi-frequency fabrication on the same substrate particularly challenging [17] [14].

More recently, Lithium Niobate devices leveraging Lamb waves on thin-film plates have attracted interest. Laterally Vibrating Resonators (LVRs) offer high electromechanical coupling (30 − 40%) and a wide frequency range on the same substrate (2-16 GHz), offering significant flexibility in manufacturing and integration [18] [19]. Nonetheless, their dispersive behavior is strictly related to the thickness of the piezoelectric film, and they require careful energy confinement to minimize acoustic losses with quality factor ($Q$) and electromechanical coupling coefficient ($k_t^2$) co-optimized as a function of dispersion, particularly at higher frequencies [18] [20].

Laterally excited Bulk Acoustic Wave (LBAW) resonators, initially explored by Plessky et al. (2019) as XBAR [21], and further investigated by Villanueva et al. (2024) as B-IDT [22], represent a significant breakthrough. By embedding electrodes within the piezoelectric medium, these resonators efficiently excite bulk acoustic waves laterally, for both $S_0$ and $SH_0$ modes. Compared to LVRs, LBAWs exhibit significantly enhanced static capacitance density ($C_0$), lower series resistance ($R_s$) and reduced ohmic losses, improved power handling and heat dissipation, as well as the possibility to exploit viable overtones with substantially relaxed lithographic constraints. Compared to LVRs — whose dispersion characteristics are highly sensitive to film thickness — LBAWs offer enhanced on-chip tunability and scalability: for any desired frequency range, the optimal configuration can be achieved simply by adjusting the electrode-to-piezoelectric width ratio, independently from film thickness.

## III. Design and Optimization

Device optimization is carried out for an X-cut Lithium Niobate thin-film with thickness equal to 100 nm. Ideally, the electrodes should span through the entire thickness, however this would greatly complicate fabrication efforts. The electrodes depth is therefore set to 80% of the piezoelectric thickness, i.e. 80 nm. Targeting an operational frequency around 1 GHz for the fundamental $S_0$ mode, the piezoelectric width ($w_p$) is fixed at $2\mu m$. By means of Finite Element Analysis (FEA) carried out with the aid of COMSOL® Multiphysics, dispersion curves for varying metal-to-piezoelectric width ratios ($w_m/w_p$) are simulated, identifying optimum working conditions for both $S_0$ and $SH_0$ fundamental modes, and their first odd overtones.

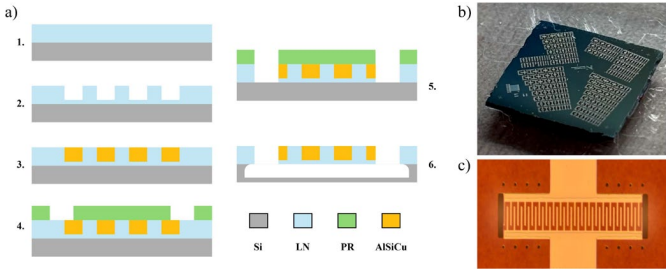

Fig. 5. a) Fabrication flow schematic summary as detailed in Section IV. b) Fabricated chip image highlighting different device orientations. c) Optical microscope image of unreleased resonator after performing release pits etching with self-alignment method.

Similarly to FBAR, minimal electrode widths are ideal to excite the fundamental tone, concentrating the stress in the piezoelectric region [17]. Specifically, simulations indicate peak $k_t^2$ values of 28% for $S_0$ modes at $w_m/w_p = 0.1$ and up to 43% for $SH_0$ modes at $w_m/w_p = 0.1125$. Conversely, LOBAW devices display similar behavior to that observed in Overmoded Bulk Acoustic Resonators (OBARs), necessitating large electrode widths to accommodate the additional semi-wavelengths of the second overtone [23] [24]. The optimum points are identified at $w_m/w_p = 1$ for the $S_0$ overtone, and $w_m/w_p = 1.1$ for the $SH_0$ overtone, with simulated coupling coefficients of $k_t^2 = 13\%$ and $k_t^2 = 21\%$, respectively (Fig.2 and Fig.3).

To conclude this investigation, the energy distribution at resonance is evaluated. By computing the energy confinement ratio $\eta$, it is possible to estimate acoustic losses, particularly the higher the energy density in the electrodes, the higher the losses [25]. The obtained curves show that the $w_m/w_p$ values that optimize coupling coefficient are very close to local maxima of the energy confinement ratio; as both quantities are optimized for the same geometrical configuration, no net trade off between coupling and quality factor is to be expected (Fig. 4).

Once again, it is worth noticing that the performed optimization is not sensitive to the thickness of the piezoelectric layer, but only to the geometrical dimension of the interdigitated structure. By decoupling the thickness from the dispersion curves, it is possible to optimize any target frequency on the same chip by simply adjusting the $w_m/w_p$ ratio.

## IV. Fabrication and Characterization

Fabrication is carried out on a 10 mm chip of X-Cut LN 100 nm thick thin-film bonded on High Resistivity Silicon provided by NGK Insulators, Ltd. Electron beam lithography is used to define the interdigitated structures. Resonators are patterned along crystallographic propagation directions identified to maximize the electromechanical coupling coefficient for each mode under investigation. On X-Cut LN, $S_0$ and $SH_0$ modes are optimized for Euler angles equal to $(-90°, -90°, 30°)$ and $(-90°, -90°, -10°)$ respectively. Using the same mask, electrodes are initially etched via timed Deep Oxide Etching (DOE), subsequently 80 nm of AlSiCu metal are deposited by thermal evaporation to metalize the trenches. Using the self-align method, described in [26], released windows are etched again via DOE. Similarly to LVRs [18], LBAWs require half-width electrodes on the edge of the plate to impose the correct boundary condition; by patterning the full electrode and aligning the etch pit so that half of the metal is removed, it is possible to further relax lithography constraints. After depositing 300 nm thick AlSiCu pads for $G-S-G$ probing, the plates are released via isotropic $XeF_2$ etch.

Upon release, the devices are characterized via direct RF probing in a 2-port configuration with the aid of Vector Network Analyzer (VNA). Measurements are carried out in high vacuum (5 mTorr). The measured $S$-parameters are then converted in admittance $Y_{12}$ via software and the resonator's response is fitted via single tone modified Butterworth-Van Dyke (mBVD) model [27].

The best observed performances (shown in Fig. 6 and Fig .7) are found for $w_m/w_p = 0.075$ for the fundamental tones and $w_m/w_p = 1.1$ for the overtones, showcasing Figure of Merits (FoM) of 437 at 673 MHz for the $SH_0$ fundamental tone and 53 at 1.05 GHz for the $SH_0$ overtone. The low quality factor at antiresonance $Q_p$ might be attributed to improper release of the wide plates. Additionally, parasitic series capacitance is introduced by the oxidation of the AlSiCu in the overlapping region between probing pads and buses.

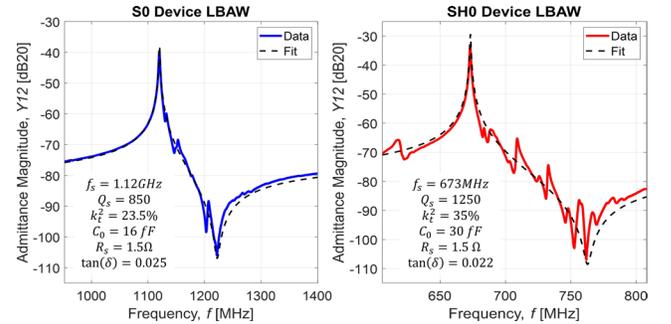

Fig. 6. Extracted $Y_{12}$ admittance response and corresponding equivalent mBVD fitting parameters for $S_0$ LBAW (left) and $SH_0$ LBAW (right).

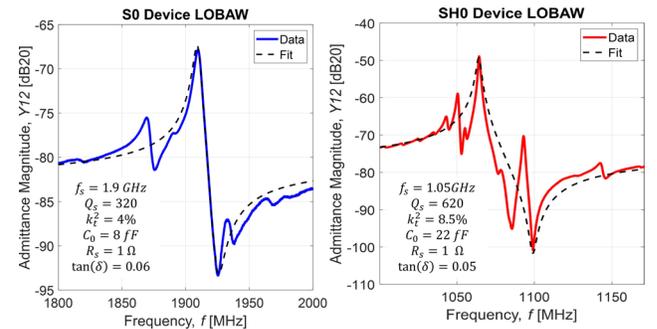

Fig. 7. Extracted $Y_{12}$ admittance response and corresponding equivalent mBVD fitting parameters for $S_0$ LOBAW (left) and $SH_0$ LOBAW (right).

## V. Conclusions and Remarks

In this work, Laterally Excited Bulk Acoustic Wave (LBAW) resonators on X-cut Lithium Niobate were designed, fabricated, and characterized, with Figures of Merit of 437 at 673 MHz and 53 at 1.05 GHz for the $SH_0$ fundamental mode and overtone respectively, marking the first ever demonstration of viable overtones for this type of resonator.

Embedding electrodes into the piezoelectric layer allows to decouple the dispersion relation from the film thickness, thus enabling on-chip multi-frequency scalability and tunability. While leaving room for improvement in the design and fabrication process, for instance to reduce the impact of flexural spurious modes and improve the quality factor, the obtained results are encouraging towards the possible deployment of this architecture in telecommunication systems as a viable alternative to traditional SAW/BAW technologies.

## VI. Acknowledgements

Authors Walter Gubinelli and Luca Colombo equally contributed to this work. The authors would also like to thank Northeastern Kostas Cleanroom and Harvard CNS staff.